\documentclass[pre,aps,twocolumn,showpacs,floatfix]{revtex4-1}

\usepackage{dcolumn}
\usepackage{amsmath}
\usepackage{amssymb}
\usepackage{bm}
\usepackage{graphicx}
\usepackage{epsf,epsfig}
\usepackage{hyperref}
\usepackage{array}
\usepackage{tabularx}
\usepackage{xcolor}

\begin{document}

\title{The role of vimentin-nuclear interactions in persistent cell motility\\ through confined spaces}
\author{Sarthak Gupta$^{1}$, Alison E. Patteson$^{1}$, J. M. Schwarz$^{1,2}$}
\affiliation{$^1$ {Physics Department and BioInspired Institute, Syracuse University, Syracuse, NY USA}, \\
$^2$ Indian Creek Farm, Ithaca, NY USA}
\date{\today}
\begin{abstract}
The ability of cells to move through small spaces depends on the mechanical properties of the cellular cytoskeleton and on nuclear deformability. In mammalian cells, the cytoskeleton is composed of three interacting, semi-flexible polymer networks: actin, microtubules, and intermediate filaments (IF). Recent experiments of mouse embryonic fibroblasts with and without vimentin have shown that the IF vimentin plays a role in confined cell motility. Here, we develop a minimal model of a cell moving through a microchannel that incorporates explicit effects of actin and vimentin and implicit effects of microtubules. Specifically, the model consists of a cell with an actomyosin cortex and a deformable cell nucleus and mechanical linkages between the two. By decreasing the amount of vimentin, we find that the cell speed increases for vimentin-null cells compared to cells with vimentin. The loss of vimentin increases nuclear deformation and alters nuclear positioning in the cell. Assuming nuclear positioning is a read-out for cell polarity, we propose a new polarity mechanism which couples cell directional motion with cytoskeletal strength and nuclear positioning and captures the abnormally persistent motion of vimentin-null cells, as observed in experiments. The enhanced persistence indicates that the vimentin-null cells are more controlled by the confinement and so less autonomous, relying more heavily on external cues than their wild-type counterparts.  Our modeling results present a quantitative interpretation for recent experiments and have implications for understanding the role of vimentin in the epithelial-mesenchymal transition.

\end{abstract}               
\maketitle


\footnotetext{\textit{$^{*}$~E-mail: sgupta14@syr.edu}}
\footnotetext{\textit{$^{**}$~E-mail: aepatte02@syr.edu}}
\footnotetext{\textit{$^{***}$~E-mail: jschwarz@physics.syr.edu}}

\section{Introduction}
Cell migration is a fundamental process that contributes to building and maintaining tissue. To be able to migrate, the cell cytoskeleton, which is comprised of three dynamic polymeric systems: F-actin, microtubules, and intermediate filaments (IFs), generates forces.  While actin and microtubules are more studied cytoskeletal filaments, intermediate filaments (IFs) also play a role in a range of cell and tissue functions ~\cite{Leduc2015, Danielsson2018, Pekny2007}. Vimentin is an IF protein whose expression correlates with {\it in vivo} cell motility~\cite{Helfand2011, Mendez2010a} behaviors involved in wound healing~\cite{Eckes2000, Rogel2011a} and cancer metastasis~\cite{Kidd2014, Satelli2011}, and, yet, its role in three-dimensional cell migration is poorly understood.

{\it In vivo}, cells move through a confining tissue environment made out of other cells and extracellular matrix (see. Figs 1(a) and 1(b))~\cite{Friedl2000}. Emerging experimental studies show that the structure and dynamics of the cytoskeleton in three-dimensional motility differ from those for cells on surfaces~\cite{Friedl2000, Petrie2012}.  In confined settings, actin tends to accumulate at the cell cortex~\cite{Balzer2012b} and microtubules align with the direction of the confining track~\cite{Doyle2009}. Moreover, in highly confining environments, the nucleus, one of the stiffest organelles in the cell, can be an inhibitor of cell migration due to steric hindrance~\cite{Friedl2011}. Earlier cell motility models in such environments, therefore, focus on the nucleus~\cite{Reversat2020, LeBerre2013, Hawkins2009}, in contrast with cell motility models for surfaces~\cite{Danuser2013}.  Recent studies have shown that in three-dimensional settings the centrosome displays an increased probability to be near the rear of the cell~\cite{Zhang2017}. When cells change direction in a narrow track, the centrosome repositions, moving from one side of the cell to the other, to repolarize the cell by developing a new trailing edge of the cell and setting up the polar direction of the cell. Interestingly, the nucleus appears to be decoupled from this phenomenon in that the removal of the cell nucleus does not alter the centrosome repositioning in most cells~\cite{Zhang2017}. Though for strong confinement, the cell nucleus will presumably play a more dominant role in the repositioning of the centrosome as it deforms. 

\begin{figure*}[t]
\includegraphics{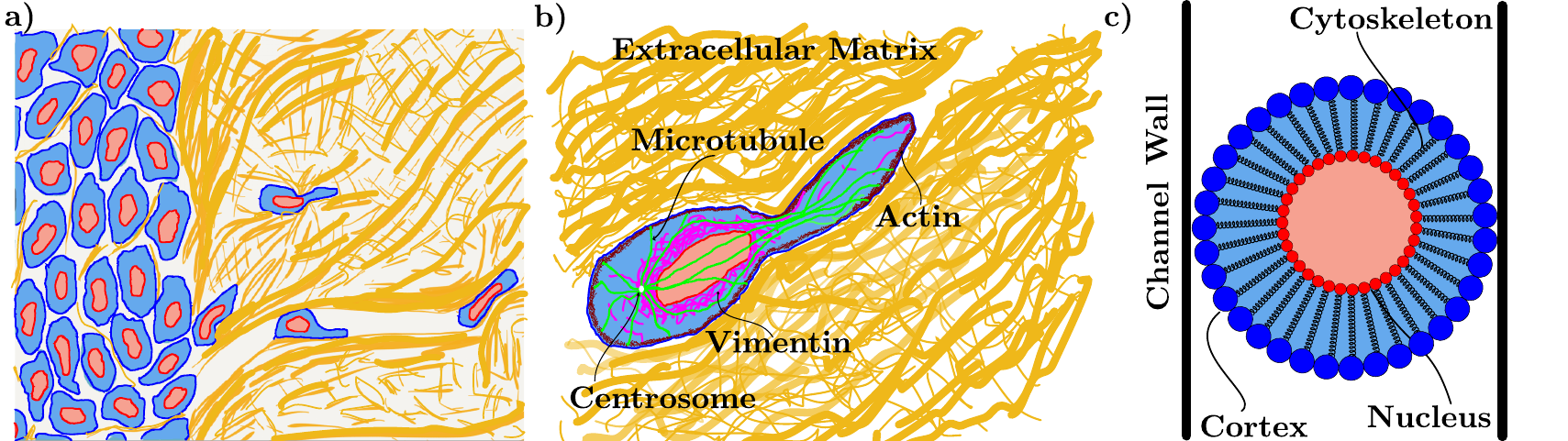} 
  \caption{{\it Modeling confined cell motility:} (a) Cells move through confining environments due to the extracellular matrix (in gold) and other cells. (b) Schematic of an individual cell with a cytoskeletal network containing actin, microtubules, and intermediate filaments, a nucleus, and a centrosome connected to the nucleus via the protein dynein. (c) A simulation model for the cell: The cortex is made up of blue monomers connected with springs, the nucleus is also made up of red monomers with springs connecting them. The bulk cytoskeletal network is simplified and modeled as springs connecting the nucleus and the cortex. The cell and nucleus contain cytoplasmic and nucleoplasmic material, each of which is modeled as area springs.
}
\end{figure*}

Given the roles of actin, microtubules, and centrosome positioning in confined cell motility, one is naturally led to ask about the role of vimentin, which couples to both actin and microtubules and forms a cage around the cell nucleus~\cite{Costigliola2017}. Recent experiments~\cite{Patteson2019f, Battaglia2018} highlight new roles of vimentin in mediating cell speed and cell persistence in three-dimensional confining environments.
Specifically, when comparing the motion of wild-type mouse embryonic fibroblasts (mEFs) with their vimentin-null counterparts moving through three-dimensional micro-fluidic channels, the loss of vimentin increases cell motility and increases rates of nuclear damage in the form of nuclear envelope rupture and DNA breaks. Moreover, unlike in unconfined motility, the loss of vimentin increases the spontaneous persistent migration of cells through 3D micro-channels and 3D collagen matrices \cite{VanBodegraven2020}.

Inspired by these experiments, we seek to quantitatively understand the role of vimentin in confined cell motility via a computational model. To do so, we develop a minimal, yet detailed, cell motility model incorporating actin, microtubules, and vimentin. The starting point of the model is the notion that vimentin plays a distinct role in mediating forces between the actomyosin cortex and the nucleus~\cite{Patteson2020, Patteson2021}.  Our model, therefore, contains both an actomyosin cortex and a nucleus, whose interaction via a set of linker springs is strengthened by the presence of vimentin. With this level of detail, a new mechanism for cell polarity for motility in which vimentin plays an important role naturally emerges. As the cell moves through the model micro-channel, we will quantitatively show how vimentin modulates cell speed, nuclear shape, dynamics, and cell persistence. In addition to quantitative comparison with experiments, we pose new insights for the role of vimentin in confined cell motility more generally. In particular, we address the upregulation of vimentin typically found in the epithelial-mesenchymal transition~\cite{Mendez2010a} in terms of how vimentin affects a cell's interaction with its environment.

\section{Model}

\subsection{The players}
{\it Cellular cytoskeleton:} The cellular cytoskeleton is composed of actin, microtubules, and intermediate filaments. The persistence length of actin filaments is smaller than microtubules, but larger than intermediate filaments ~\cite{Janmey1991b}. Myosin motors exert forces on actin filaments to reconfigure them. Many actin filaments and myosin motors reside in proximity to the cell membrane to form the actomyosin cortex.  The actomyosin cortex is an important piece of the cell motility machinery as its reconfiguring drives cell motion. Microtubules typically originate from the microtubule-organizing center, or centrosome, and have a crucial role in cell polarity as they as required to generate traction forces~\cite{Bouchet2017}.  Microtubules also have a role in controlling cell shape as they typically span the entire cell and are the stiffest cytoskeletal filaments ~\cite{Janmey1991b}. Vimentin filaments exist as a cage or mesh structure around the nucleus and are also present in the cytoskeleton in fibrous form as can be seen in Fig. 1(b)~\cite{Costigliola2017, Patteson2020}. Studies show that vimentin provides structural integrity to the cell~\cite{Patteson2019, Guo2013}. 

The various types of cytoskeletal filaments do not act independently of each other. In fact, actin, microtubules, and intermediate filaments are rather interconnected with each other~\cite{Seetharaman2020}. For instance, vimentin and actin directly interact with each other via the tail domain of vimentin~\cite{Esue2006a}. Moreover, plectin is a major crosslinker among all three types of filaments~\cite{Svitkina1998}. 

{\it Cell nucleus:} The nucleus is typically the largest, and stiffest, organelle in the cell, yet it is still deformable. Recent studies showed a cell under a highly confined environment, and thereby, the nucleus is so squeezed such that DNA becomes damaged~\cite{Patteson2019}. Cells cannot move through a particular confining geometry if the nucleus cannot do so~\cite{Friedl2011}. Therefore, the nucleus is also an important player in confined cell motility. There also exists a  nuclear envelope, or nuclear cortex, consisting of inner nuclear lamins and outer vimentin.

The interconnected cytoskeletal networks and the nucleus are also connected to each other. For instance, LINC(Linker of nucleoskeleton and cytoskeleton) complexes made of nesprins and SUN (Sad1 and UNC84) proteins act as connecting bridges between nucleoskeleton, containing lamins, and the actin network in cytoplasm~\cite{Crisp2006}. Microtubules are also joined to the nucleus via kinesin-1 which talks to nesprin-4 \cite{Roux2009a}. Intermediate filaments are connected via plectin which connects to nesprin-3, which is joined with nucleus~\cite{Wilhelmsen2005b}. Disruption of these nucleus-cytoskeleton links also leads to impairment of 3D cell migration~\cite{Khatau2012, Harada2014}. 

{\it Centrosome:} Mechanisms for driving cell polarity, or the direction in which a cell moves, is also crucial for understanding cell motility. Recent experiments in two-dimensional motility suggest that removing the centrosome induces microtubules to grow symmetrically in all directions and, as a result, the cell forms lamellae in many directions and, therefore, loses its polarization~\cite{Wakida2010}. In a 3D setting, the centrosome is typically found to be posterior of the cell, and microtubules are oriented in the direction of motion (for certain cell types)~\cite{Doyle2009}.  Therefore, the position of the centrosome sets the polarity of the cell by defining the tail/back of the cell during migration since the change in cell direction results only after the centrosome moves to the new posterior side of the cell~\cite{Ueda1997}.  

\subsection{Cell as a spring network}
Given the complexity of such interactions of the above players, we took a reductionist approach and simplified this highly tangled picture to a two-dimensional network of harmonic springs with an outer ring of springs representing the cellular cortex and an inner ring of springs representing the nuclear envelope, and harmonic springs connecting the cellular cortex with the nuclear cortex (see Fig. 1(c)). Our model is a two-dimensional cross-section of the three-dimensional system. As we can see from the schematic, each cell cortex monomer is joined to each nuclear cortex monomer via a linker spring.  Each spring type, the cellular cortex, the nuclear cortex, and the interacting bulk linker spring, has its own stiffness with the respective potential energies $V_{cc}=\frac{K_{cc}}{2} (r_{cc}-r_{cc,o})^2$,  $V_{nc}=\frac{K_{nc}}{2} (r_{nc}-r_{nc,o})^2$, $V_{l}=\frac{K_{l}}{2} (r_{l}-r_{l,o})^2$, respectively, with $r_{cc}$ denoting the distance between the centers of neighboring cell cortex monomers, for example, and $r_{cc,o}$ represents the rest length of the spring. We also include potentials in the form of two area springs, $V_{cell}=\frac{K_{cell}^{area}}{2} (A_c-A_{c,o})^2$ and $V_{nuc}=\frac{K_{nuc}^{area}}{2} (A_n-A_{n,o})^2$, where $A_{c}$ and $A_n$ denote the areas of the cell and nucleus, respectively. The two area springs prevent the cell and the nucleus from collapsing as the cell becomes increasingly more confined. 

How do we explore the role of vimentin in such a mechanical model, particularly when considering wild-type fibroblasts versus their vimentin-null counterparts? Since vimentin-null cells are softer than wild-type cells~\cite{Guo2013} and exhibit more DNA damage~\cite{Patteson2019}, to capture both cell lines we change the stiffness of the cytoplasmic/linker springs $\text{K}_{\text{l}}$ and nucleus area springs $\text{K}_{\text{nuc}}$. Since removing vimentin does not significantly affect the cortical stiffness~\cite{Guo2013}, we do not alter the cortical spring constant among two cell lines. 

The cell also interacts with its confining walls which are considered to be adhesive due to fibronectin or collagen I or some other kind of protein that is usually coated inside the channels. Specifically, the micro-channel is modeled as two lines of wall monomers fixed in place that interact with the cell cortex monomers via adhesion. In previous studies, adhesion to the surface is modeled as catch and slip bonds~\cite{Bangasser2013a, Lopez2014}, where the cell forms a bond to the substrate, and then as it moves those connections peel off. We, therefore, model the adhesion interaction with the Weeks-Chandler-Anderson potential, or

\begin{equation}\label{eq2}
   V_{cell-wall}=
    \begin{cases}
      4\epsilon_{Ad} \left [ \left ( \frac{\sigma_{cw}}{r} \right )^{12} - \left ( \frac{\sigma_{cw}}{r} \right )^{6} \right ] + \epsilon & r\leqslant 1.2\sigma_{cw}\\
      0 &  r> 1.2\sigma_{cw},
    \end{cases}       
\end{equation}
where $\epsilon$ quantifies the adhesion strength and the minimum of the potential is located at $2^{1/6}\sigma_{cw}$. While the potential is purely repulsive in the short-range, there is also an attractive component. Cell experiences no adhesion until the distance between cortex and wall is less than or equal to $1.2\sigma_{cw}$. After which, cell experiences an attractive force till the potential minima $2^{1/6}\sigma_{cw}=1.122\sigma_{cw}$. The difference between cutoff distance and potential minima $0.077\sigma_{cw}$ or $0.15507\, \mu m$ or $155.07\, nm$, thus represents a typical focal adhesion size ~\cite{Changede2017a, Evans2007}.  We did not vary the adhesion strength between the two cell lines.

\subsection{Polymerization and adhesion forces}
 As the cell moves, actin is polymerized at the leading edge of the cell to translate the cell in a particular direction with microtubules setting the direction~\cite{Meiring2020}. We model actin polymerization via an active force, $\mathbf{F}_a$. The active force is present for half cortex of the cell, the leading edge half, and has a magnitude $F_a$. The direction of $\mathbf{F}_a$, which is set by microtubules emanating from the centrosome, is initially chosen to be towards the opening of the micro-channel, which determines the leading edge half---its polarization direction. There are small fluctuations in the direction of $\mathbf{F}_a$ as it moves through the micro-channel. 

{\it Polarity-mechanism:} If one allows for large fluctuations in the direction of $\mathbf{F}_a$, the probability is enhanced of the cell turning around in the channel. Cell length is also found to be correlated with the time required for cells to change the direction, which is the time the centrosome takes to move to the other side of the cell to define a new tail~\cite{Zhang2017}. Therefore, we posit that the more the centrosome is located away from the cell center of a crawling cell, the more biased cell migration becomes. Thus, the centrosome is essential for the preservation of polarized cell morphology. 

While there is no explicit centrosome, nor microtubules, in our model, the centrosome is also connected to the nuclear outer membrane by a protein emerin~\cite{Salpingidou2007}. Given the strong coupling between the nucleus and the centrosome, we effectively include a centrosome and a direction of cell polarity as determined by the direction of microtubule polymerization.  To do so, we define $\mathbf{d}$ representing the difference between the center of mass of the cell and the center of mass of the nucleus. Its angle is measured from the positive $x$-axis. To be specific, if the cell moves parallel to the channel away from the designated entrance, $\theta=90^o$; if the cell performs a ``180'' to head directly back towards the designated entrance, $\theta=180^o$. For incorporating the centrosome role in motility, we define the following empirical equation:

 \begin{equation}\label{eq3}
\Delta \theta =\frac{\pi}{2}\left ( 1-\frac{|\mathbf{d}|}{R_{nuc}} \right ),       
\end{equation}
where $R_{nuc}$ denotes the rest radius of the nucleus. With this empirical equation, we define the bounds on the direction a cell can take. We take the nuclear cortex-cellular cortex (NC-CC) axis as our reference axis and for the upper bound and lower bound we can add and subtract $\Delta \theta $ from this axis. See Fig. 4(a). Then, we choose a random angle under these bounds and that represents the direction of the nuclear, or centrosome, axis given the strong coupling between the two, which then drives the direction of migration. 

We found that the negative $\Delta \theta$ is highly unlikely in this model as the purely repulsive LJ interaction ensures that the distance between the center of the cell and the center of the nucleus does not increase more than the radius of the nucleus. For instance, when the nucleus gets displaced off-center, the maximum it can go is to the cortex wall. Also, even if $\Delta \theta$ is negative, the upper and lower bound of available angles would simply flip. Thus, it would not change the range of available angles for choosing the next direction. In our current analysis, we ignore the ensembles where $\Delta \theta$ is negative.

What is the timescale for changing $\Delta \theta$? We assume that vimentin acts as a template for microtubules in the cell on a time scale of 10-20 minutes since (1) vimentin has a slow turnover rate than microtubules and (2) microtubules in vimentin-null cells show less orientation than in wild-type cells~\cite{Gan2016}. Therefore, this polarity mechanism repeats itself every $15$ minutes as, on average, this is the time after which vimentin restructures ~\cite{Gan2016, Yoon1998}. For simplicity, we assume that the repolarization time in the vimentin-null cell is the same as wild-type, as it is not known if microtubule turn-over time changes between the two or not. We leave the investigation of this assumption for future work. 

Additionally, we ensure that all cells move toward the channel initially so that they will be in contact with the mouth of the channel. Then the polarity mechanism kicks in and the cell chooses a new direction. We have used the initial radius of the nucleus as a normalization constant in Eq. 3. Note that there is no memory of the previously taken direction. The NC-CC axis is, again, taken as a reference after each $15$ min. and depending on the distance of the nucleus from the cell cortex,  $\Delta \theta $ is calculated, and eventually, a new angle is again chosen randomly from within the bounds. Note that the cell does not exactly follow the NC-CC axis for cell migration due to the intracellular dynamics and fluctuations in the cell.

In addition to adhering to the wall, it has been observed that in confined environment experiments, actin bundles start forming at the edges of the cell where it is interacting with the walls~\cite{Paul2016}. We are proposing that due to this interaction with the wall, the cell generates forces to enhance motility in the direction in which it is polarized. Therefore, any cell cortex monomer with some proximity of a wall monomer exerts an additional force, $\mathbf{F}_w$ in the direction of the leading edge. For avoiding overlap (volume exclusion), we apply $\mathbf{F}_{rep}$ which is derived by WCA potential with a cut-off at minima($2^{1/6}\sigma$), which makes this force purely repulsive in short-range and zero for a greater distance than minima. $\mathbf{F}_{rep}$ is applied to all particles except particles connected via two-body harmonic spring.
\begin{figure*}[t]
\includegraphics{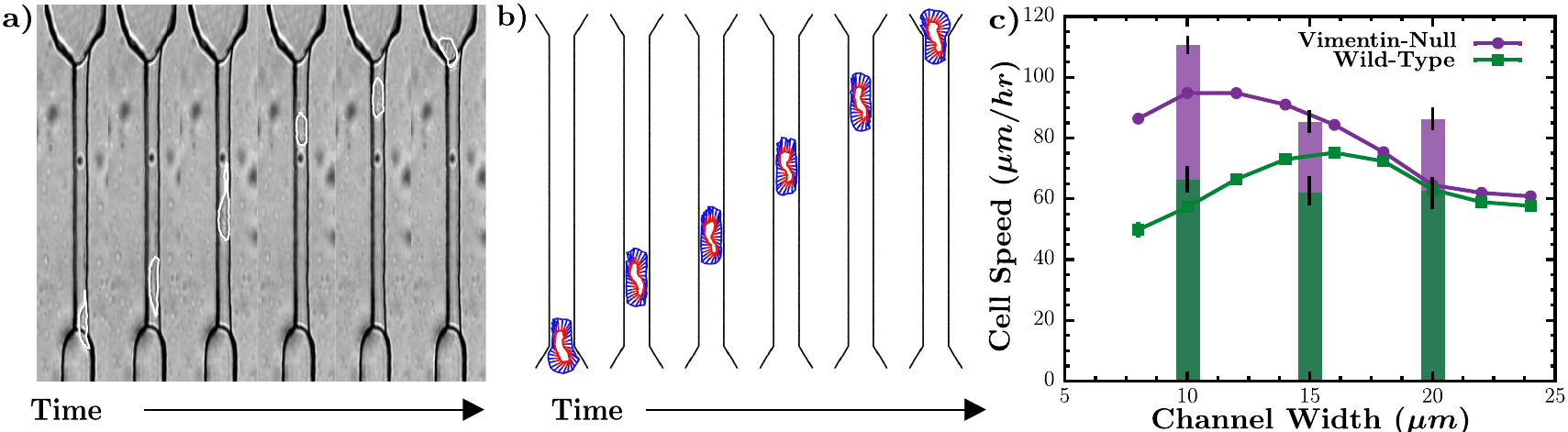}
  \caption{ {\it Model quantitatively agrees with cell speeds from micro-channel experiments}:  (a) Time series of a wild-type mEF cell moving in a microchannel (b) Same as (a) but images are from a computational model. (c) Average cell speed as a function of channel width for wild-type and vimentin-null cells for the computational model. Prior experimental results are also plotted for comparison.  Note that the experimental results indicate the average maximum cell speed, while the computational results indicate the average speed. }
\end{figure*}
\subsection{Dynamics}
Now that we have detailed the forces involved, here is the equation of motion for each cell monomer of type $i$ at position $\mathbf{r}_i$: 
\begin{equation} \label{eq1}
\dot{\mathbf{r}}_i=\mu_i (\mathbf{F}_{a} +\mathbf{F}_{w} +\mathbf{F}^{c}) + \sqrt{2D_{i}} \mathbf{\xi}_i(t),
\end{equation}
 where $\mathbf{F}_{a}$ is an active force representing the actin forces at the front end of the cell in the direction of motion and $\mathbf{F}_{w}$ denotes the force generated at the wall in the direction of the leading edge. Finally, $\mathbf{F}^{c}$ represents the conservative forces in the system, or  $\mathbf{F}^{c}=\mathbf{F}_{cc}+\mathbf{F}_{nc}+\mathbf{F}_{l}+\mathbf{F}_{cell}+\mathbf{F}_{nuc}+\mathbf{F}_{cell-wall}+\mathbf{F}_{rep})$, which are the two-body springs, the area springs, both modeling the mechanics of the cell and the adhesion force between the cell and the confinement (wall) and purely repulsive volume-exclusion force. For the parameters varied in the simulations, see Table 1.
 
Regarding parameters, the diffusion constant of both the cell and nuclear membranes are rooted in biologically relevant time scales such that membrane displacements are much less than the global motion of the cell and nucleus, respectively. The diffusion constant of the cell membrane, $D_{cc}$, is estimated from the actin bundle motion just below the cell membrane \cite{Shirai2017a}. For estimating the nuclear membrane diffusion constant, $D_{nc}$, we look at the fluctuations of the cell nucleus~\cite{Chu2017}. The values $D_{cc}$ and  $D_{nc}$ are close, but not the same ($D_{cc}>D_{nc}$ ) to account for the fact that these two objects reside in slightly different environments. See Table 1. As for the mobilities, prior work has implemented cell cortex mobilities an order of magnitude lower~\cite{Lopez2014}.  There is some range of applicability of these estimates as they are indeed estimates. A different mobility can be absorbed into changing the time scale and the strength of the noise (see Eq. 2).

We use simulation units defined as unit simulation length equal to $\mu \text{m}$, unit simulation time is equal to $sec$ and unit simulation force is $nN$. The respective diameter of the monomers are $\sigma_{cc}=\sigma_{cw}=2\, \mu m$, $\sigma_{nc}=1 \,\mu m$, all neighbouring springs in actomyosin cortex and nuclear cortex is $2\, \mu m$ and $1\, \mu m$ respectively and linker spring length is $5.73 \, \mu m$. Relevant length scale for converting kPA to nN/$\mu m$ are cortex diameter($20\, \mu m$), nucleus diameter($10\, \mu m$), and linker spring length for cortex spring, nucleus spring, and linker spring strength respectively. Our simulations used $36$ monomers in both cortices, $74$ monomers to simulate each side of the straight part of the wall, and $7$ monomers for each side of the slanted channel entry and exit. To iterate Eq. 2, we use the Euler-Maruyama method. \\

\begin{table}[h!]
\centering

\begin{tabularx}{0.5\textwidth} { c  c  c  c}
 \hline\\
 Parameters & Wild-type & Vimentin-null & Refs.  \\\\
\hline
\hline\\
 $K_{cc}$  & $100\, nN/ \mu m$ & $100 \,nN/ \mu m$ & \cite{Guo2013}  \\\\
 $K_{nc}$  & $1000\, nN/ \mu m$ & $1000\, nN/ \mu m$   & \cite{Caille2002, Vahabikashi2018} \\\\
 $K_l$  & $10\, nN/ \mu m$ & $1\, nN/ \mu m$  & \cite{Guo2013, Vahabikashi2018}\\\\
 $K_{cell}^{area}$  & $0.01 \,nN/ \mu m^{3}$ & $0.01 \,nN/ \mu m^{3}$  & -  \\\\
 $K_{nuc}^{area}$  & $0.05 \,nN/ \mu m^{3}$ & $0.005 \,nN/ \mu m^{3}$   & -\\\\
 $F_a$  & 	$3\, nN$ 			& $3 \,nN$   & \cite{Prentice-Mott2013}\\\\
 $F_w$  & $3\, nN$ & $3 \,nN$   & -\\\\
 $\epsilon_{ad}$  & $1 \,nN\, \mu m$ & $1\, nN \, \mu m$ & -\\\\
 $\epsilon_{rep}$  & $1 \,nN\, \mu m$ & $1\, nN \, \mu m$ & -\\\\
 $D_{cc}$  & $0.02\, \mu m^{2}/s$ & $0.02\, \mu m^{2}/s$ & \cite{Shirai2017a}  \\\\
 $D_{nc}$  & $0.04\, \mu m^{2}/s$ & $0.04\, \mu m^{2}/s$ & \cite{Chu2017}  \\\\  
 $\mu_{cc}$  & $0.01\, \mu m/nN\cdot s$ & $0.01\, \mu m/nN\cdot s$ & -  \\\\
 $\mu_{nc}$  & $0.02\, \mu m/nN\cdot s$ & $0.02\, \mu m/nN\cdot s$ & -  \\\\

 \hline
\end{tabularx}
\caption{Table of parameters used, unless otherwise specified.}
\label{table:1}
\end{table}

\begin{figure*}[t]
\includegraphics{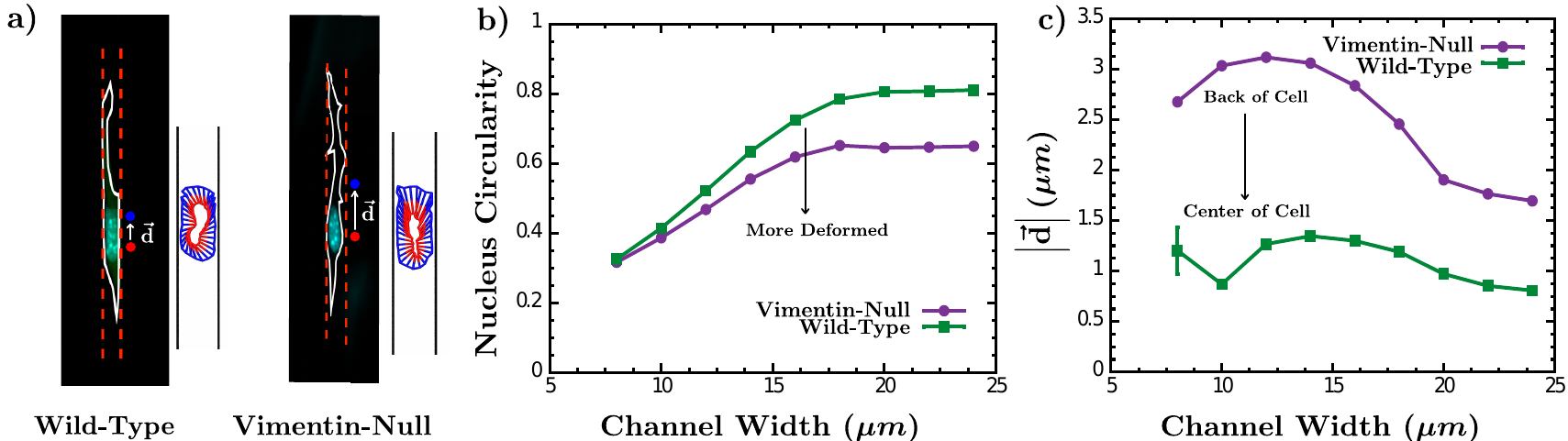}
  \caption{{\it Vimentin affects nuclear shape and position}. (a) Images of a wild-type mEF cell and a vimentin-null mEF cell moving in a microchannel and their computational counterparts. In addition, the vector difference between the cell center of mass and the nucleus center of mass, $\vec{d}$, is labeled on each cell image. (b) The circularity of the cell nucleus as a function of confinement for both cell types. (c) The position of the center of mass of the cell with respect to the center of mass of the nucleus as a function of channel width for both cell types.} 
\end{figure*}  
\subsection{Model summary} We model the cell as containing an outer actomyosin cortex, represented by monomers connected by springs, a nuclear cortex, also represented by monomers connected by springs, and bulk cytoskeletal filaments---including vimentin, represented as linker springs connecting the two inner and outer cortices.  See Fig. 1(c). There is an energetic cost to changing the area of both cortices as well to capture the incompressibility of the cell and the nucleus. The cell also interacts with each wall of the microchannel composed of fixed monomers via an adhesive potential. To study the motility of the cell through the microchannel, there is an active force, $\bf{F}_a$ on the half of the acto-myosin cortex monomers in a direction initially towards the opening of the microchannel and whose direction is determined by the microtubule-centrosome system.  Should any actomyosin cortex monomer come within some proximity of the wall, there is an additional force, $\bf{F}_w$, exerts on an actomyosin cortex monomer in the direction of the leading edge.  The direction of the leading edge is chosen uniformly at random within some range of angle.  We propose a form for the range of possible leading edges that depends on the magnitude of the difference between the center of the mass of the cell and the center of mass of the nucleus, with the position of the nucleus serving as a readout for the position of the centrosome. The forces in the model are updated using over-damped dynamics.  For a detailed explanation motivating each force in the model, please see the Model section above.  See Figs. 2a and 2b for snapshots of the experiments and the simulations of a cell moving through a microchannel. Two supplementary movies generated with Visual Molecular Dynamics (VMD) \cite{Humphrey1996} have also been provided.

\subsection{Analysis}
 We measure the average speed of the center of mass of the cell while it is in the channel. This cell speed average is then averaged over approximately 1000 realizations for each channel width. Error bars represent the standard deviation of the mean of approximately $10^{3}$ ensemble runs. We also measure the persistence in the motility as defined as the ratio of the path length of the center of mass of the cell divided by the length of the channel.  Flux is defined as the fraction of cells that exit the channel on the side different from the entry side. The circularity of the nucleus $C$ is defined as  $C=\frac{4 \pi A}{P^2}$, where $A$ is the area of the cell and $P$ is its perimeter. When $C=1$, the nucleus is a circle. The time-averaged energy of the cell is calculated by taking the ensemble average of energy contributions from the conservative potentials. The initial cell energy for the conservative potentials is zero.

\section{Results}

\subsection{Cell speed is non-monotonic with confinement and affected by vimentin}

Strikingly, we observe for both cell types that as the channel width decreases, the average cell speed is non-monotonic (Fig. 2c). This trend has also been observed in experiments \cite{Charras2014,Irimia2009b,Jacobelli2010,Balzer2012}. How does such a trend emerge? As the channels become narrower, the cell's cortex increases its contact with the wall. This increase in contact generates more driving force to increase cell speed. However, this trend is competing with the deformability of the cell.  As the channel width becomes even narrower, the linker springs become more deformed (more compressed) and so these springs act to increase the effective adhesion to the wall given that unbinding to the wall is driven by a distance threshold. This increased adhesion time leads to a slower cell speed.  Thus, the non-monotonic trend emerges from these competing factors, contact with the wall increasing the driving force and increasing the adhesion time as the cells become increasingly more deformed.  We can modulate this competition by either increasing the driving force or the adhesion strength. If one increases the driving force, either by increasing the active force or the wall force, not only will the maximum speed increase, the peak will broaden towards smaller channel widths (Fig. S1). If one decreases the adhesion strength, a similar effect occurs (Fig. S2).

How is such a trend modified in vimentin-null cells? Since the presence of vimentin makes cells stiffer~\cite{Eckes1998, Wang2002, Guo2013, Patteson2019f}, the vimentin-null cell line is described by a decreased linker spring strength and a decreased nucleus area spring strength, with the latter capturing the lack of a vimentin cage around the nucleus. We indeed observe a similar non-monotonic trend, but with a larger average cell speed as compared to wild-type cells, at least for narrower channel widths. Moreover, the maximum average cell speed occurs at a narrow channel width. Again, as confinement increases, the bulk cytoskeleton in both cell lines also starts deforming (compressing). However, the wild-type cell provides more resistance as it is a stiffer cell, and, thus, pushes against the walls more to effectively act a  stronger adhesion to the wall. This effective adhesion to the wall is weaker for the vimentin-null cell line because linker spring strength is weaker.  This effect results in both a larger maximum average cell speed and the driving force out-competing the adhesion for a larger range of change of channel widths. Note that our results also depend on the nucleus area spring strength. Decreasing the nucleus area spring strength also decreases the effective adhesion to the wall, as the anchoring of the springs to the nucleus is less stiff and so also enhances the average cell speed (Fig. S3).

Now let us compare the computational model with the experiments. Our average cell speed is in 
reasonable quantitative agreement with the experimental cell speed measurements for both cell types and for two different channel widths (Fig. 2(c)) with wild-type cells moving more slowly than vimentin-null cells in the microchannels channels~\cite{Patteson2019f}. This outcome is in stark contrast with two-dimensional cell motility studies ~\cite{Helfand2011}. The experiments demonstrate that as confinement increases, vimentin-deficient cell's average speed also increases, whereas wild-type cell speed remains largely unchanged~\cite{Patteson2019f}. We find similar behavior with the non-monotonic trend in cell speed as a function of channel width weaker in the wild-type case. This non-monotonicity also agrees with the experimental observations of other cell lines migrating in the channels~\cite{Charras2014}. For vimentin-null cells, we predict that for channels less than a 10-micron width that the cell speed will begin to decrease.  Hints of this prediction are evident in transwell experiments~\cite{Patteson2019}.

\subsection{Vimentin's dual role of stress transmitter and nuclear protector}

 \begin{figure*}[t]
\includegraphics{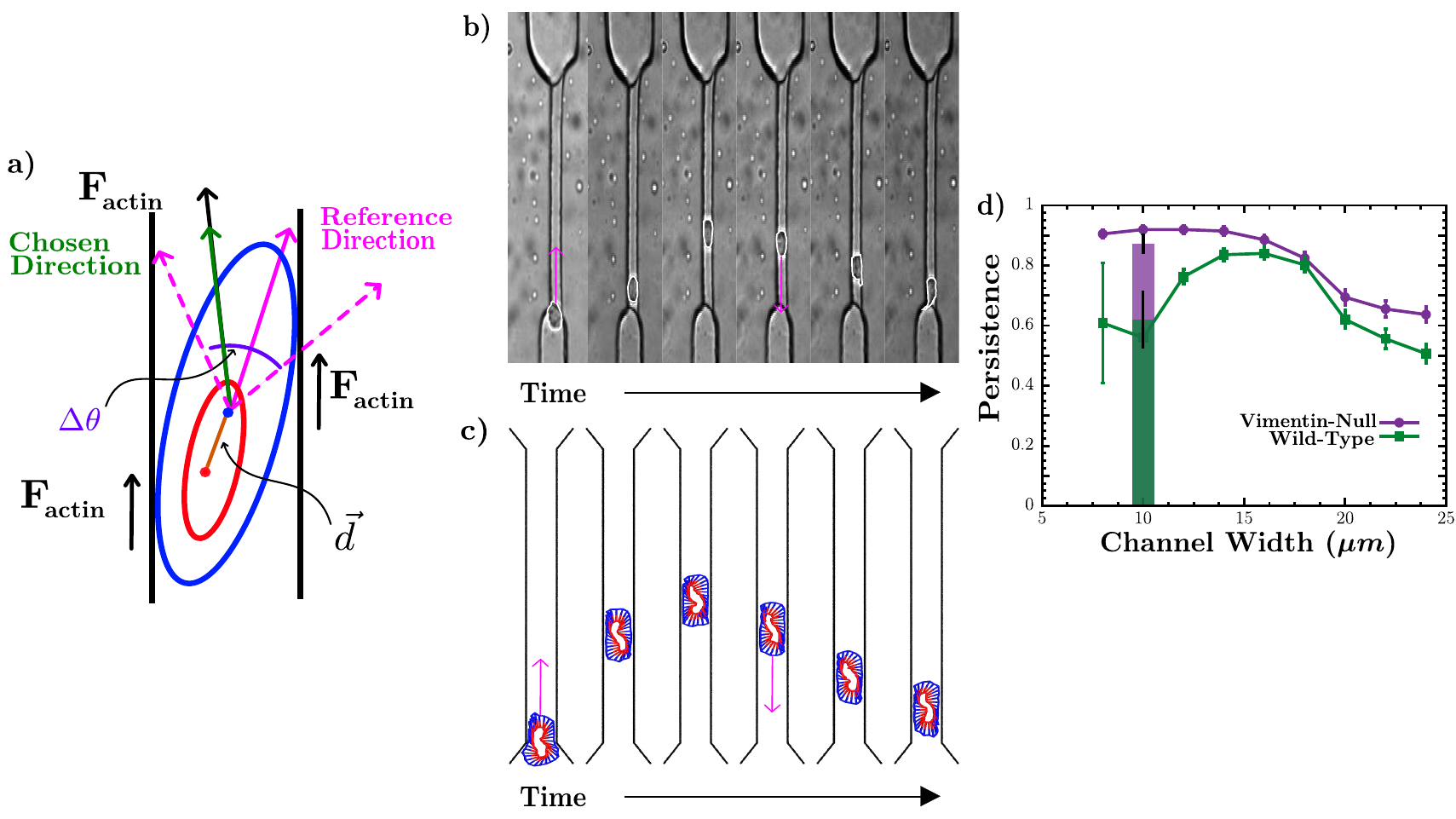} 
  \caption{{\it Nucleus-centrosome-based polarity mechanism}. (a) Schematic of the polarity mechanism. (b) Time series of a cell moving into the channel and changing direction. For experimental methods, please see \cite{Patteson2019f}  (c) Computational example of a cell changing direction in the channel. (d) Pers istence as a function of channel width for both cell types.} 
\end{figure*}

Since the linker spring strength mediates the transmission of forces between the actin cortex and the nuclear cortex, one, therefore, may anticipate the shape of the nucleus to depend on the linker spring strength. For instance, we expect the nucleus shape to be less correlated with the cell cortex shape as the linker spring strength becomes weaker. Prior experiments indeed indicate that the shape of the nucleus is affected by vimentin as cells move in confinement~\cite{Lowery2015}. To quantify the effects of the linker spring strength on nuclear shape, we calculate nuclear circularity $\text{C}$, which is defined as $\frac{4 \pi A}{P^{2}}$, where $A$ denotes the cross-sectional area of the nucleus and $P$ its cross-sectional perimeter (Figs. 3(a) and (b)). When $\text{C}=1$, the nuclear cross-section is a circle. As the channel width decreases, nuclear circularity decreases for both cell types, however, the nuclei of the vimentin-null cells are typically more deformed than their wild-type counterparts. The greater decrease in $C$ for the wild-type cells with increasing confinement is due to the larger stresses mediated by the bulk cytoskeleton, thereby potentially motivating the need for a mesh-like vimentin cage around the nucleus to help mediate them. 

Interestingly, experiments indicate that the nucleus in vimentin-null cells has more wrinkles than wild-type and has less effective volume~\cite{Lele2018}. As discussed in the modeling section, in addition to modifying the linker spring strength to go from one cell type to the other, we also modify the area spring constant for the nucleus, with the latter accounting for the mesh-like vimentin cage surrounding the nucleus. When we increase the area spring constant for the nucleus, as well as increase the linker spring stiffness, we find that even though the cytoskeleton still transfers forces from outside to the nucleus with increasing confinement, the nucleus resists such deformations with the circularity tracking very similarly for the smaller channel widths between the two cell types. To see the further effects of linker spring constant and nuclear area spring constant on nucleus shape, please refer to Fig. S10.

The nuclear envelope is typically viewed as dominated by the nuclear laminas~\cite{McGregor2016}. Here we suggest that vimentin around the nucleus also has a role in stabilizing the shape of the nucleus. Thus, we can conclude that vimentin protects the nucleus from deformations as well as mediates stress transmission between the two cortices, the actin one and the nuclear one~\cite{Maniotis1997, Neelam2015}.  Now we have quantitative modeling results to substantiate this concept. 

Since nuclear positioning is important for cell migration, we next compute the center of mass of the nucleus, the center of mass of the cell, and determined the distance between the two, defined as the magnitude of $\vec{d}$ with the origin of $\vec{d}$ at the center of mass of the nucleus as shown in Figs. 3(a) and 3(c). This metric estimates the location of the nucleus with respect to the rest of the cell.  When $|\vec{d}|\approx 1 \mu m$, the nucleus is closer to the center of the cell as compared to the vimentin-null case.  For larger $|\vec{d}|$, the nucleus moves toward the rear of the cell or away from the leading edge of the cell. We find that the nuclei in the vimentin-null cells are typically positioned toward the rear of the cell, while the nuclei for the wild-type cells are closer to the center of the cell (Fig. 3(c)). As the channel width narrows, this difference becomes even starker.  

How does this trend of nuclear positioning emerge?  For the more malleable linker springs, springs in the leading half of the cell more readily attach to the wall creating a typically flatter leading edge for the part of the cell cortex not attached to the walls.  The linker springs in the leading half of the cell associated with the leading edge are, therefore, more extended (see Fig. 3(a)).  From an energetic point of view, the extra tension (stretching) in the leading half of the cell is compensated for by the springs in the rear half of the cells configuring to be close to their rest length.  This argument also tells us that the nucleus is being pulled by the leading half of the cell as it moves through the channel.  We confirm this by calculating the forces on the nucleus due to the cortex (Fig. S4). Our results hopefully prompt additional experiments focusing on nuclear positioning for confined cell motility.

\begin{figure*}[t]
\includegraphics{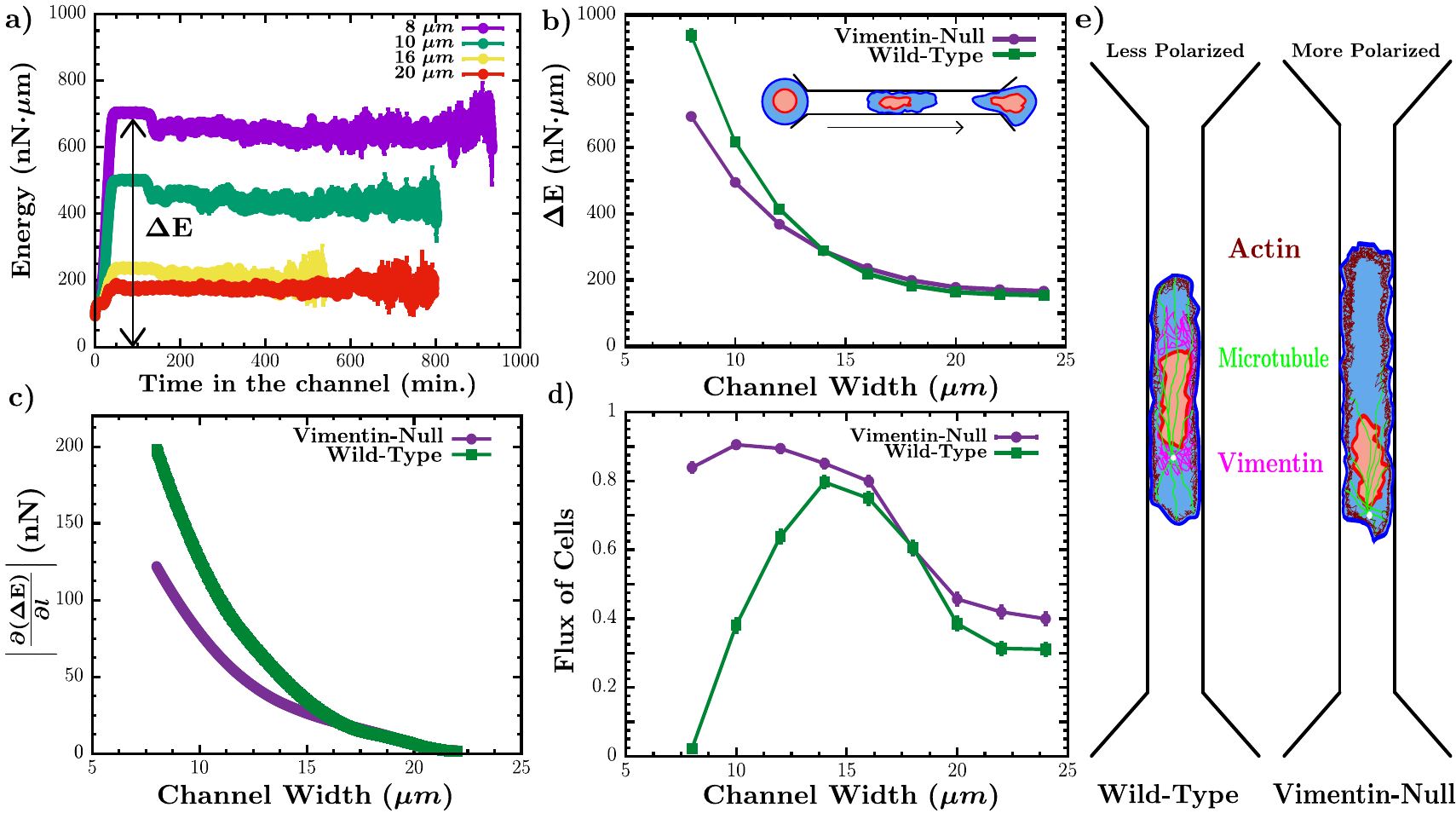} 
  \caption{{\it Energy barriers in confined cell motility}. (a) The average energy of each cell due to the conservative forces as a function of time for different channel widths. (b) The time-averaged energy as a function of channel width for both cell types. (c) The magnitude of the numerical derivative of the time-averaged energy, which is proportional to the stress on the cell due to the channel, as a function of channel width for both cell types. (d) The flux of each cell type as a function of the channel width. (e) A schematic of the internal organization of more polarized/persistent cells versus less polarized/persistent ones with the centrosome denoted by the white circle.} 
\end{figure*}

\subsection{Loss of vimentin increases cell persistence in micro-channels}

Based on the observed changes in nuclear positioning, we propose a novel polarity mechanism based on the position of the nucleus, which is a readout for the position of the centrosome in the cell. The centrosome plays an important role in cell polarity and is also connected to the nucleus via various crosslinkers and proteins \cite{Luxton2011}. We assume that such connections between the centrosome and nucleus remain strong as the cells move through the channel and so the position of the nucleus tracks the position of the centrosome.  We posit in confinement that as the magnitude of $\vec{\text{d}}$ increases, the cell becomes more polarized with the possible range of angles of microtubule polymerization scaling as $|\Delta \theta|= \frac{\pi}{2} \left (1-\frac{\left | \vec{d} \right |}{R_{Nuc}}  \right )$, where $R_{Nuc}$ is the rest radius of the nucleus and $\pm\Delta\theta$ is defined clockwise/counter-clockwise from the reference angle of $\vec{d}$ respectively (Fig. 4(a)). There are two reasons for this scaling.  First, as the nucleus is more displaced, only the longest microtubules polymerize towards the leading edge of the cell. The lateral confinement and rear end of the cell disrupt microtubule polymerization in the remaining directions. This spatial arrangement biases microtubule polymerization parallel to the confining walls ~\cite{Lagomarsino2007, Pinot2009}.  See Supplementary Fig. S5 for the distribution of angles possible for the leading edge of the cell (with respect to the $x$-axis) and the distribution of angles chosen by the cell.

In Figs. 4(b) and (c), we show the time series for a wild-type cell from the earlier micro-channel experiment~\cite{Patteson2019f} and compare it with our computational model. 
As in the experiments, we find that the wild-type cells are more likely to change direction in the channel. To quantify this, we measure the contour length of the trajectory normalized by the length of the micro-channel such that a persistence of unity occurs when the cell does not change direction in the channel. We find that the vimentin-null cell line is more persistent in the channels as compared to the wild-type for all channel widths but more notably different for the smaller channel widths (Fig. 4(d)). We have also studied the persistence as a function of $k_L$, $k_{Nuc}^{Area}$, $F_a$, $F_w$, and $\epsilon_{Ad}$. See Supplementary Figs. S6, S7, S8. Given the more asymmetric nuclear positioning in the vimentin-null, a stronger $F_a$ will enhance the tensioning in the leading half of the cell and, thus, decrease $\Delta \theta$ to enhance the polarization, for example.

In the wild-type cells, we note that persistence is initially increasing with increased confinement (Fig. 4(d)). We also see there is a decreased persistence towards the tighter channels.  This behavior tracks the non-monotonic behavior observed in the magnitude of $\vec{\text{d}}$ as a function of the channel width. To explain this non-monotonic trend, we turn to the delicate balance between the stiffness of the cell and strain due to channel width. In intermediate channel widths, the wild-type spends less time, thus, has less chance of turning to the other side. The more time wild-type cells spend in the channels, the greater the chance of moving back towards the entrance or getting stuck in the channel. In wider channels, cells do not have enough wall contact but in very narrow channels, stiffness of the cells kicks in and so the cells spend more time in the channel as they travel more slowly in terms of speed. 

Indeed, our model recapitulates the experiments. More precisely, experiments found that vimentin-null mEFs were more persistent than wild-type mEFs.  For example, in $10 \mu m$ channels about half of the wild-type cells did not cross the channel, whereas most of the vimentin-null cells passed to the other side~\cite{Patteson2019f}. These results are somewhat surprising as the cells behave opposite on 2D substrates. In 2D, vimentin-deficient cells form lamellipodia in all directions thereby preventing them from polarizing, which is not as likely to occur with their wild-type counterparts.

\subsection{Energy barriers in confined migration}

It has been observed that cells tend to migrate in the direction of least confinement to minimize energetic costs~\cite{Zanotelli2019}.  To test the notion of confinement as an energetic barrier in cell motility, we compute the time-averaged energy due to conservative potentials while the cell is in the channel.  See Fig. 5(a). We repeat this measurement for the different channel widths. See Fig. 5(b). For the two cell lines, the average energy increases more for the wild-type cells than for the vimentin-null cells as the confinement increases, as anticipated.  

The energy increases as the channel width decreases and so translates to an increasing energy barrier that the cells must overcome to enter and move in the microchannel. In Fig. S9, for smaller channel widths, we demonstrate that channel width correlates linearly with cell strain.  In other words, the channel width is a placeholder for cell strain in the direction perpendicular to the walls. Since the derivative of the energy with respect to strain, or channel width, relates to the compressive stress of the cell, the compressive cellular stress increases faster than linear with decreasing channel width below some critical channel width (see Fig. 5c). In other words, the cells exhibit compression stiffening as they enter the channel.   Compression stiffening, a nonlinear rheological property in which a material's moduli increase with increasing uniaxial compressive strain, has recently been discovered in static cells~\cite{Gandikota2020}. Here, we observe a dynamic version, if you will. The vimentin-null are more deformable and so their compression stiffen is less dramatic and the onset occurs at a slightly higher strain (see Fig. 5c).

What are the implications for compression stiffening?  We observe a drop in flux for the wild-type cells as the channel width decreases.  Since flux is a measure of the fraction of those cells whose center of mass enters the channel and ultimately emerges out of the other side of the microchannel, the drop in flux indicates that a large fraction of the wild-type cells enter the microchannel but then ultimately turn around in the microchannel to come out the side they entered.  This non-linear response of the cells suggests that the cell pushes against the walls in a non-linear manner to effectively enhance the adhesion to the wall even more so than a linear material.  For the wild-type cells, the non-linear response is more pronounced.  Enhanced adhesion translates to slower speeds and, hence, more time spent in the microchannel and so there is a much higher likelihood that the wild-type cells will change direction in the microchannel to go back to the entrance of the microchannel, given the polarization mechanism at play. As discussed in the previous subsection, the persistence of the wild-type cells also decreases a consequence, again, for the smaller channel widths. 

In addition to the energetic barrier, there is also a time scale for entering the channel that translates into a rate for attempting to hop over the energy barrier. This attempt rate depends on the polarization of the cell.  Prior to entering the channel, the cell effectively sees a two-dimensional surface. In this case, the magnitude of $\vec{d}$ is smaller for both cell types, so that we anticipate more changes in cell direction for both cell types.  Since we initialize the cells to move towards the channel opening, we do not explore the attempt rate here and leave it for future work.

\section{Discussion}

To focus on vimentin's role in confined cell motility, we use a computational model that captures {\it explicitly the roles of actin and vimentin and implicitly the roles of microtubules as well as the cell nucleus}. We investigate the potential dual role of vimentin: the first being the mechanical protector of the nucleus and the second being a stress regulator between the cell cortex and the cell nucleus.  For the first role, we modify the stiffness of the nucleus, and for the second, we modify the stiffness of the mechanical connections, or linker springs, between the inner and outer cortices. For the wild-type cells, we find a nonmonotonic dependence of cell speed with channel width.   As the channel width narrows, the cell's cortex increases its contact with the wall, which, in turn,  generates more driving force to increase cell speed. Yet, this trend competes with the bulk deformability of the cell via the linker springs and nucleus to increase the effective adhesion as the channel width decreases, leading to a slower cell speed. For the vimentin-null cells, we observe a similar non-monotonic trend, but with a larger cell speed. Moreover, the maximum cell speed occurs at a narrow channel width, as compared to wild-type cells, given the enhanced bulk deformability of the cells. Thus, for nondeformable confinement with simple geometry, vimentin-null cell speed is typically faster, which is seemingly contrary to the notion that during the epithelial-to-mesenchymal-transition (EMT), epithelial cells, which are more stationary, typically upregulate vimentin to be able to move more efficiently as mesenchymal cells~\cite{Leggett2021, Lamouille2014, Eriksson2009}.

And yet, there is another ingredient to cell motility beyond the speed, which is cell direction, or cell polarity. Our results demonstrate that vimentin-null cells are more polarized in confining microchannels. This trend emerges because the nucleus is typically located more to the rear of the cell in the vimentin-null case, which biases the orientation of the longer microtubules, thereby determining the direction of the leading edge. The enhanced polarization indicates that the vimentin-null cells are more subjugated to the confinement since their own internal polarization mechanism that depends on cross-talk of the centrosome with the nucleus and other cytoskeletal connections is diminished. In other words, the {\it vimentin-null cells rely more heavily on external cues}, at least in this stiff microchannel environment, and so are less autonomous. See Fig. 5(e).  Finally, since energetic costs are known to be a predictor of the migration path in confined cell motility, we find a higher nonlinear energy barrier for wild-type cells entering more confined channels as compared to the vimentin-null cells, which, again, is seemingly contrary to the notion of the upregulation of vimentin enhancing cell motility. Moreover, restructuring of the cytoskeleton occurs on longer migration timescales to potentially alter the energy barrier.    

What do our findings tell us about the interaction between a cell and its microenvironment more generally?  In the absence of vimentin, the cells become more deformable and so more mechanically sensitive to their microenvironment. The downregulation of vimentin helps the cell travel more effectively from one place to another in a confined, straight channel. However, the real tissue environment, with its interstitial spaces, is more complex. Therefore, for a cell to upregulate vimentin in order to enhance motility translates to a cell enhancing its own internal polarization mechanism to effectively search the microenvironment for a minimal energy barrier by being able to more readily able to change direction. In other words, with the nucleus less displaced from the center of the cell, the cell is more capable of altering its own direction to be more autonomous and get itself out of potential ``dead ends'' in the extracellular environment. So cells upregulate vimentin despite the potential increase in a confinement energy barrier. They also have developed coping mechanisms, such as Arp2/3 branched actin around the cell nucleus helping it to squeeze through small pores \cite{Thiam2016} to be able to move efficiently even with the upregulation of vimentin. Such effects are not currently accounted for in the model. Additionally, with the upregulation of vimentin, there is more mechanical cross-talk between the two cortices to perhaps increase the role of the nucleus itself in regulating cell mechanics. Recent work suggests that compressed nuclei release calcium into the cytoplasm to help reconfigure the cytoskeleton \cite{Lomakin2020}.  Therefore, our theoretical findings provide a much richer interpretation of how vimentin affects cell migration with the combination of stress coupler between inner and outer parts of the cell, nuclear protector, and now polarity regulator. 

How does our model compare with other models of confined cell motility? Active gel models of the cells predict that cells push their way through confined spaces much like a climber chimneying off a wall~\cite{Hawkins2009}. Such a model does not explicitly include a nucleus nor a centrosome such that cell polarity is an input.  Another model based on the molecular-clutch mechanism explores glial cells moving through microchannels~\cite{Prahl2020}.  Cell polarity is, again, an input.  A third model with the complexity of an actomyosin cortex and a nucleus and couplings in between demonstrates that there is a nonmonotonic relationship between cell speed and matrix stiffness in two-dimensions~\cite{Pathak2018}. On the other hand, with our model, we have not incorporated actin retrograde flow as the cell is modeled as a collection of two-body and area springs.  Therefore, we do not account for such effects as actin retrograde flow. Enhanced actin retrograde flow potentially leads to faster turnover of the actin and so potentially a larger actin polymerization force as well as a faster time scale for focal adhesion disassembly that we do not account for in this model.  

Cell confinement in this model is explored by constraining the beads/nodes to stay within a channel geometry.  The cell direction of motion is randomly chosen every so many minutes. While cell speed is studied, cell persistence is not. Here, cell polarization emerges from an intra-cellular detail rooted in the position of the centrosome assumed to be in close proximity with the nucleus.  There are indeed additional models, notably, Refs. \cite{Bruckner2019a, Camley2014}. Our model walks the fine line between being minimal and yet detailed enough to quantify the new functionality of vimentin. In terms of comparison with two-dimensional cell motility models~\cite{Danuser2013},  we anticipate that the polarizability of the cell changes from our new mechanism to one that depends on the fluctuations in $d$, as opposed to the average. Moreover, we anticipate the fluctuations in $d$ being larger in the vimentin-null case, which corresponds to more possible directions in the cells.

While we have focused on comparison with mEFs, we expect our findings to generalize to other mesenchymal cell types. Specifically, another cell type would presumably translate to a different set of parameters.  Since our findings are robust for a range of several parameters, we expect our conclusions to generalize to other cell types. Given the paradigm of the EMT, we aim to test our predictions on cell types that do undergo an EMT transition while also accounting for the typical downregulation of keratin. Since we have focused here on mesenchymal cells, it would be interesting to generalize our model to include lamins in non-mesenchymal cells and explore the role of vimentin in other motility modes of migration~\cite{Petrie2015}. It would also be interesting to alter the geometry of the microchannel as well as study multiple cells moving in confinement to determine the robustness of our findings. Extending our polarization mechanism to include multicellular interactions will shed more light on the phenomenon of contact inhibition of locomotion in which motile cells stop moving or change direction upon contact with another cell.

\bibliography{library}

\section{Acknowledgements}

The authors acknowledge helpful feedback from an anonymous reviewer. JMS acknowledges NSF-DMR-1832002 and an Isaac Newton Award from the DoD. JMS and AEP acknowledge a Syracuse University CUSE grant and a Syracuse BioInspired grant. SG acknowledges a graduate fellowship from Syracuse University.

\onecolumngrid

\setcounter{equation}{0}
\setcounter{figure}{0}
\setcounter{table}{0}
\setcounter{page}{1}
\makeatletter
\renewcommand{\theequation}{S\arabic{equation}}
\renewcommand{\thefigure}{S\arabic{figure}}

\section{Supplementary Figures and Movies}
\noindent
{\it Supplementary Movie 1}: Simulation of wild-type MEFs each moving through a channel of different widths from a range of 8 to 24 microns in increments of 2 microns. \\
{\it Supplementary Movie 2}: Simulation of vimentin-null MEFs each moving through a channel of different widths from a range of 8 to 24 microns in increments of 2 microns. \\
  
\begin{figure*}[hbt!]
\includegraphics{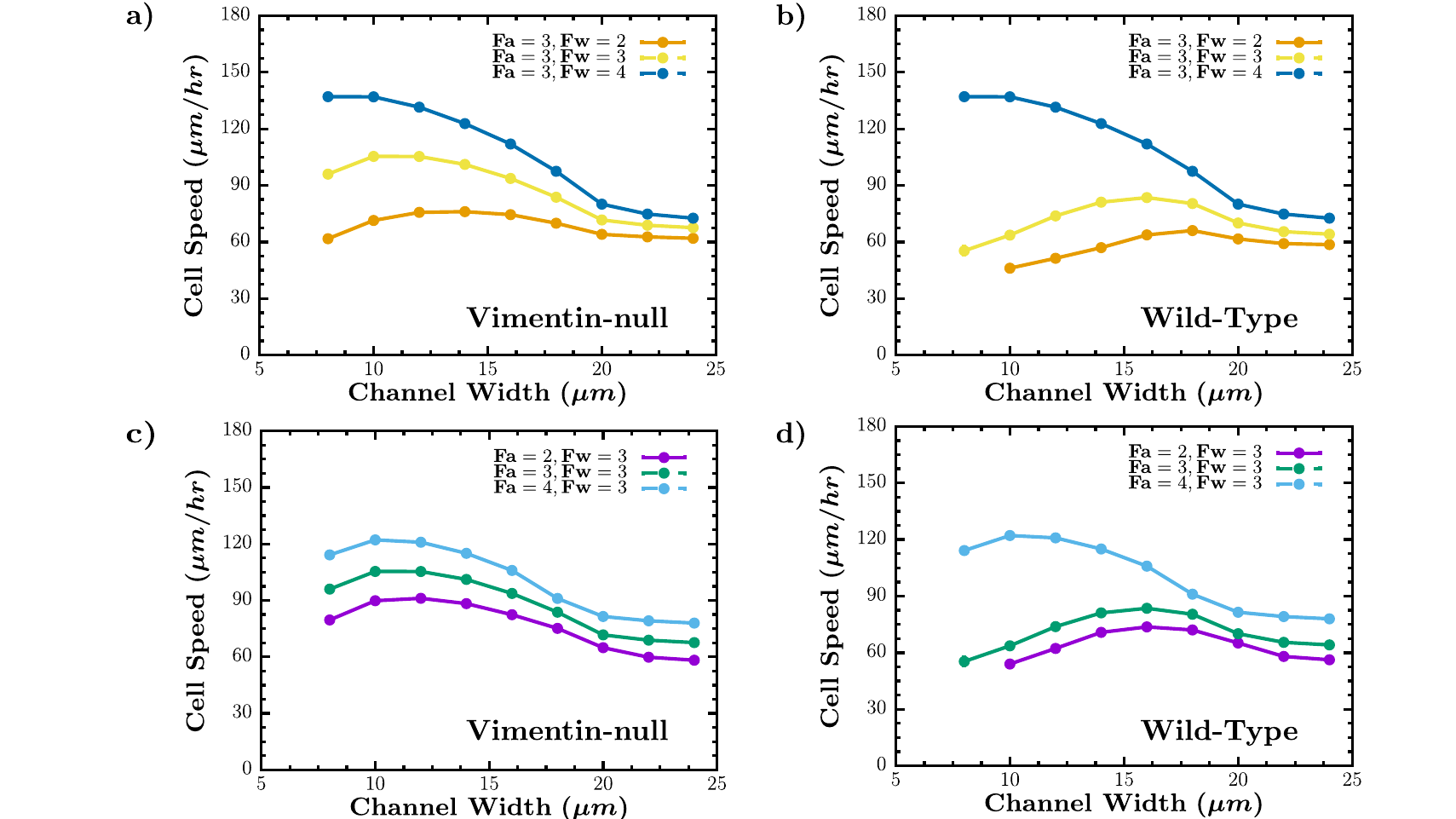}
\caption{ {\it Cell speed as a function of channel width for different actin forces at the leading edge and actin forces at the wall.'} In (a) and (b), the magnitude of the actin force at the leading edge $F_a$ is fixed and the magnitude of the actin force at the wall $F_w$ is varied for each cell type and as a function of channel width.   In (c) and (d), the magnitude of the actin force at the leading edge is now varied.} 
  \end{figure*}

\begin{figure*}[hbt!]
\includegraphics{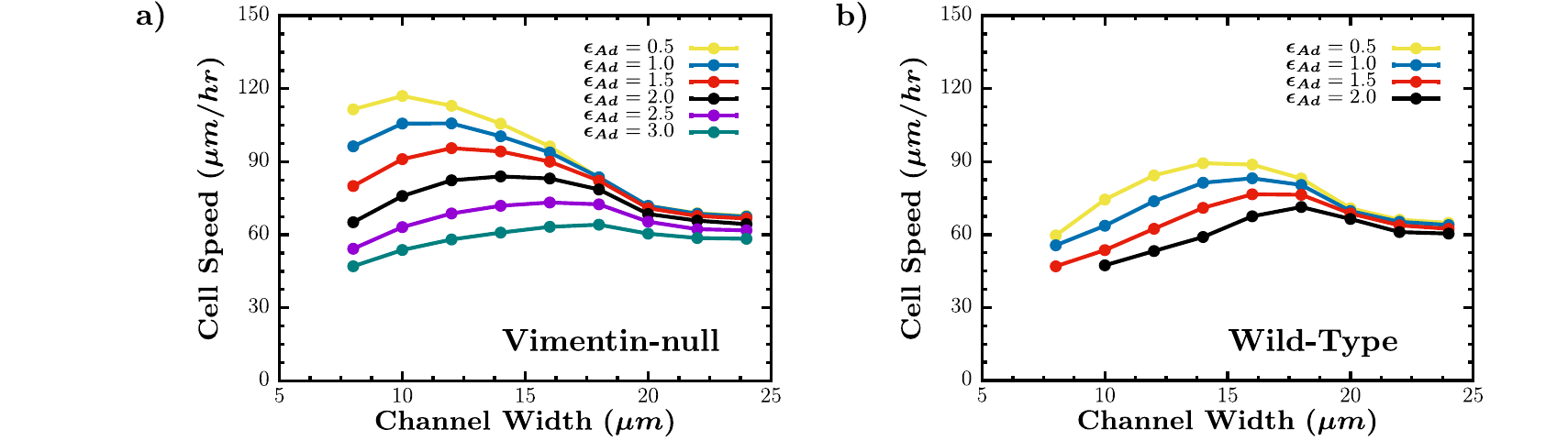}
\caption{ {\it Cell speed as a function of channel width for different adhesion strengths.} (a) Vimentin-null cell (b) Wild-type cell.} 
  \end{figure*}

\begin{figure*}[hbt!]
\includegraphics{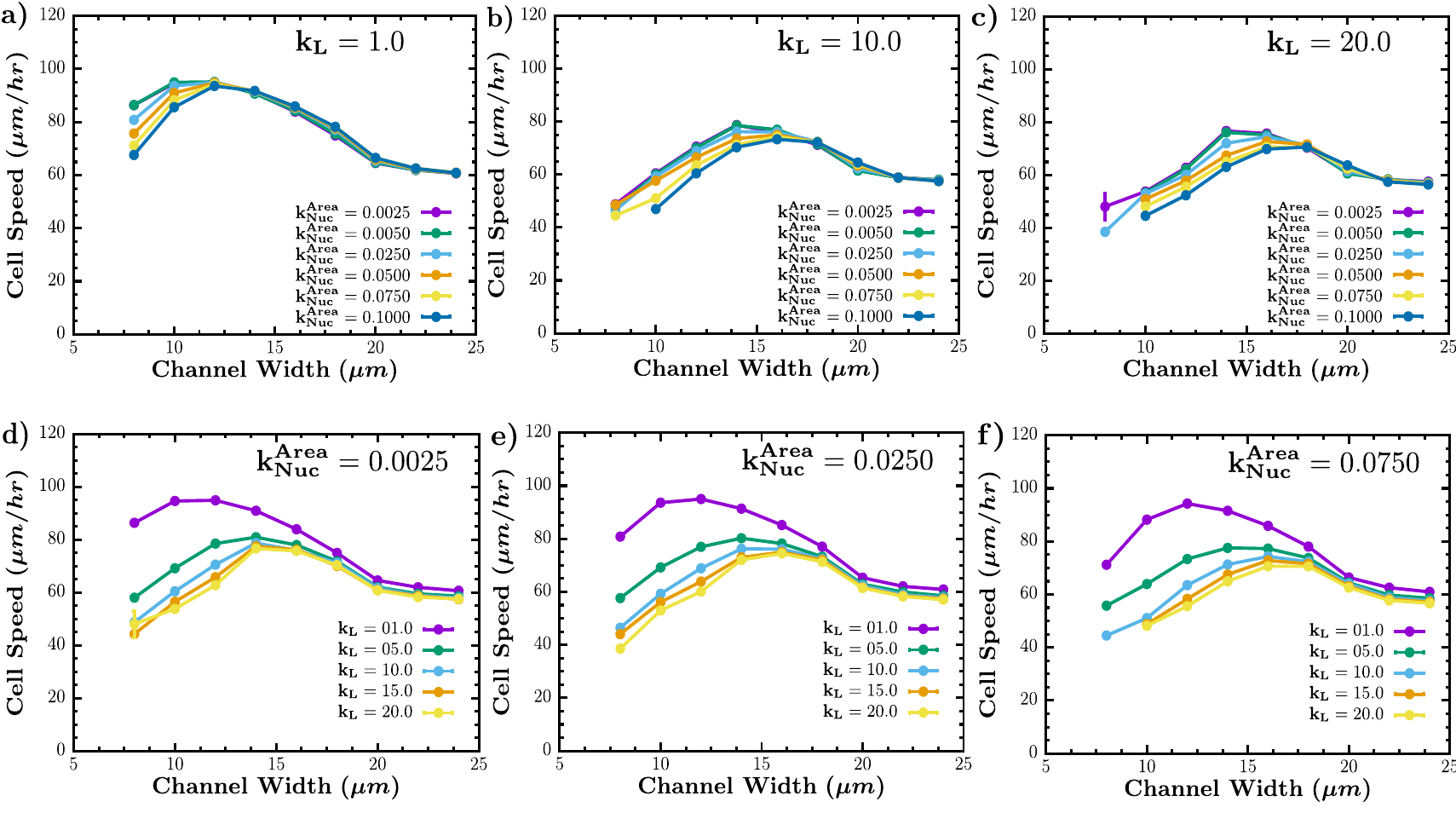}
\caption{{\it Cell Speed as a function of channel width for different linker and nuclear area spring strengths:} (a)-(c) Varying nuclear area spring strength $K_{nuc}^{area}$ for different linker spring strengths, $K_L$. (d)-(f) Varying linker spring strengths, $K_L$, for different nuclear area spring strengths  $K_{nuc}^{area}$.} 
  \end{figure*}

\begin{figure*}[hbt!]
\includegraphics{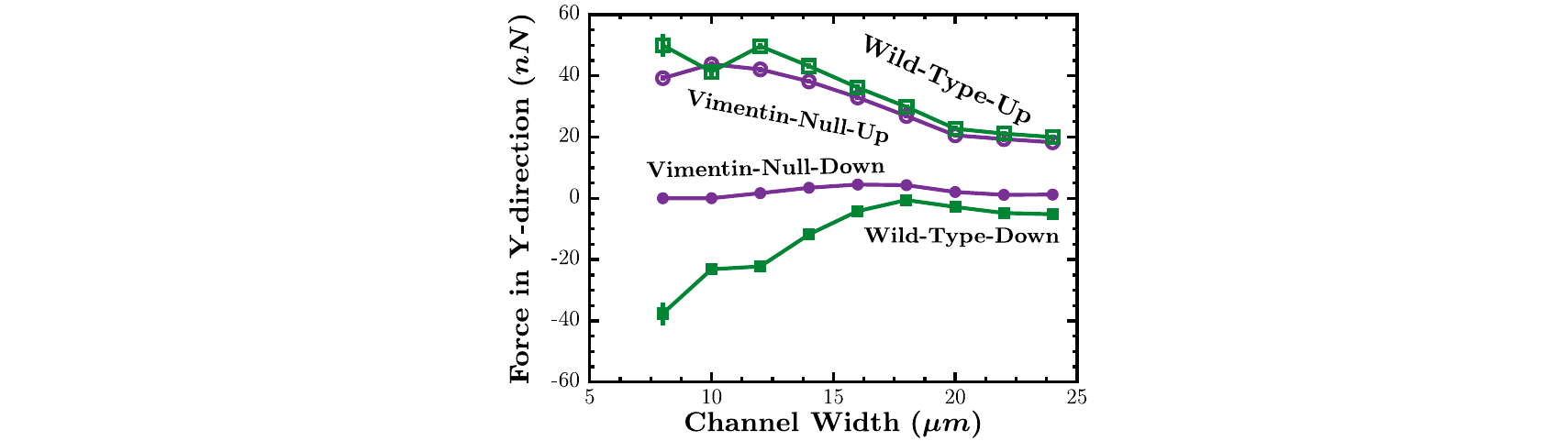}
\caption{{\it Forces on the nucleus due to the actomyosin cortex:} The net force in the y-direction on the top (upper) half of the nucleus due to the leading half of the actomyosin cortex and the net force in the y-direction on the bottom (lower) half of the nucleus due to the rear half of the actomyosin cortex as a function of channel width for each cell type. } 
  \end{figure*}

\begin{figure*}[hbt!]
\includegraphics{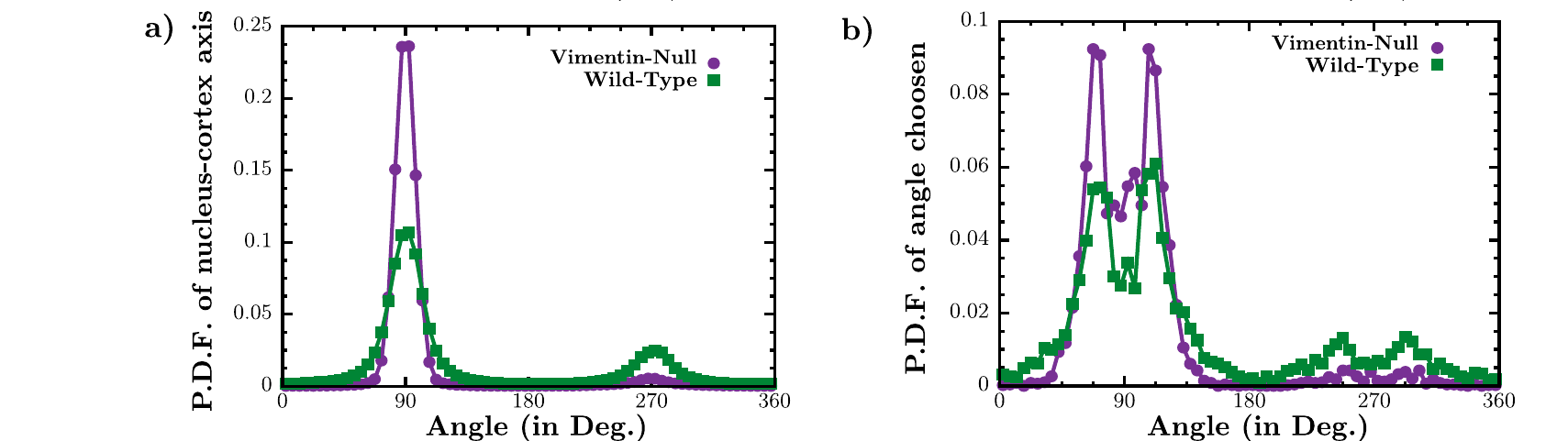}
\caption{{\it Angular information}.(a) Probability distribution function of the angle between x-axis and the Nucleus-axis for both cell types. (b) Probability distribution function of the angle choosen by the cell for a $10\, \mu\text{m}$ channel width. We are measuring all angles with respect to the $x$-axis with the nucleus center as the origin and $90^{\circ}$ as the direction towards exit from other side and $270^{\circ}$ is the direction towards the designated entrance of the channel. This figure shows multiple peaks, but the highest ones for both cell lines are around $90^{\circ}$ (exit direction or the other end of the channel) with smaller peaks at $270^{\circ}$ as well, which represent the cells turning around towards channel's entry. We observe that for vimentin-null cells, the peak around $90^{\circ}$ is higher than wild-type, which implies that the vimentin-null cells are more persistent (not turning around in the channel) than the wild-type cells.} 
  \end{figure*}

\begin{figure*}[hbt!]
\includegraphics{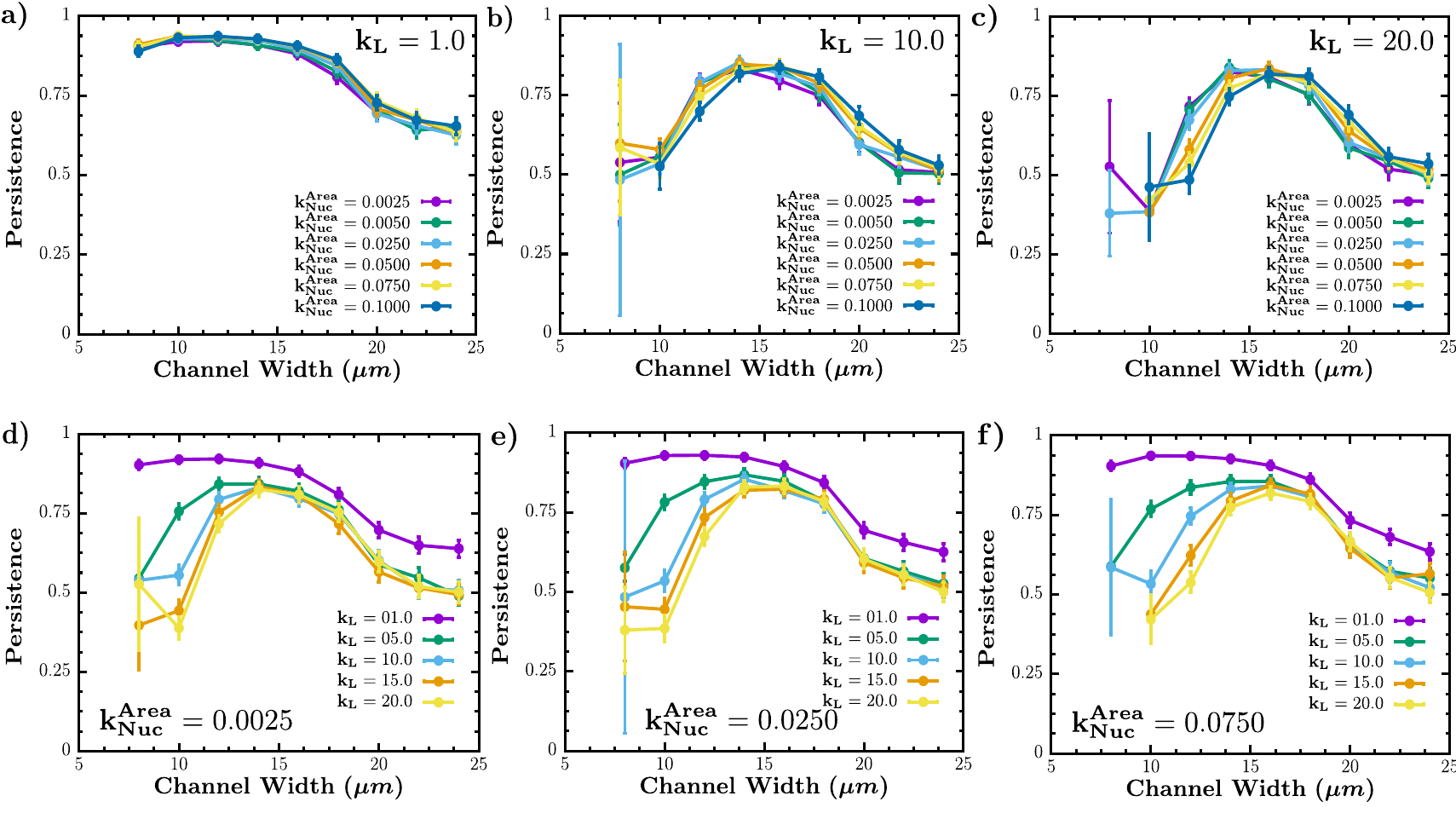}
\caption{{\it Persistence as a function of channel width for different linker and nuclear area spring strengths}: (a)-(c) Varying nuclear area spring strength $K_{nuc}^{area}$ for different linker spring strengths, $K_L$. (d)-(f) Varying linker spring strengths, $K_L$, for different nuclear area spring strengths  $K_{nuc}^{area}$.}  
  \end{figure*}

\begin{figure*}[hbt!]
\includegraphics{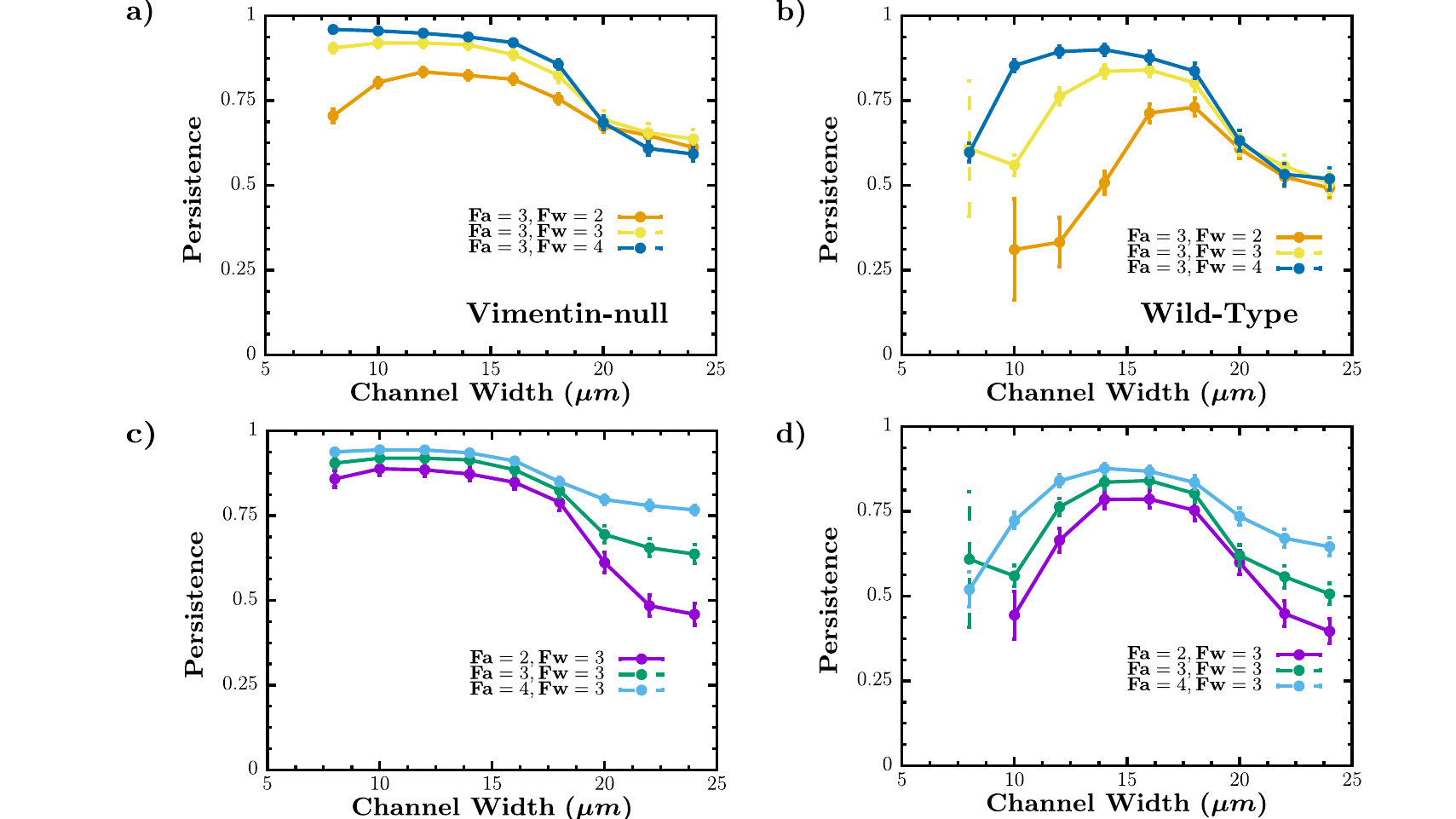}
\caption{ {\it Persistence as a function of channel width for different actin forces at the leading edge and actin forces at the wall.'} In (a) and (b), the magnitude of the actin force at the leading edge $F_a$ is fixed and the magnitude of the actin force at the wall $F_w$ is varied for each cell type and as a function of channel width.   In (c) and (d), the magnitude of the actin force at the leading edge is now varied.} 
  \end{figure*}

\begin{figure*}[hbt!]
\includegraphics{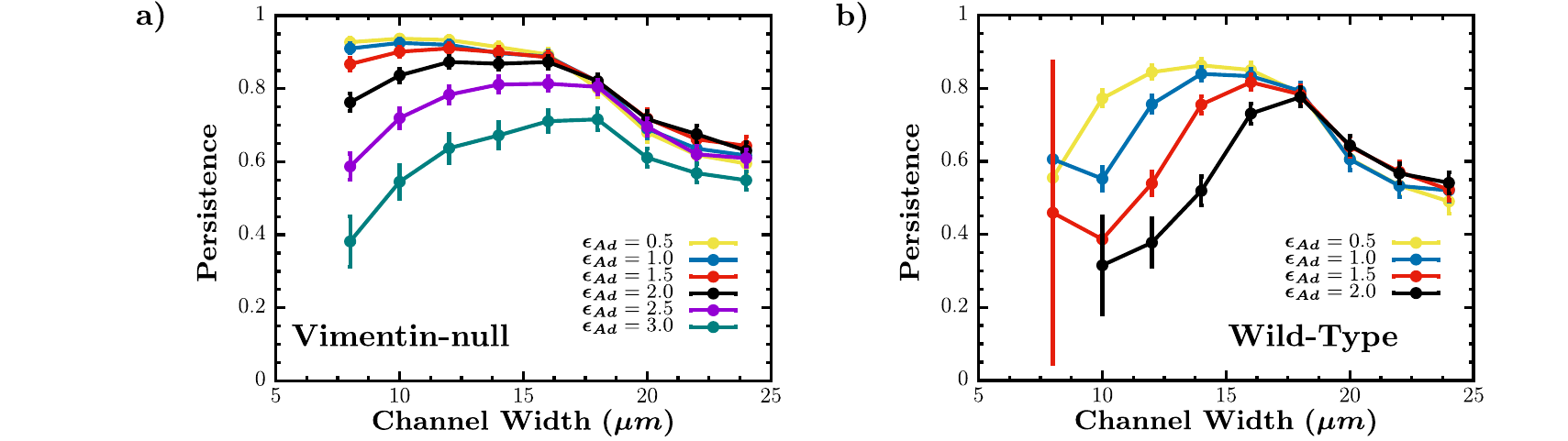}
\caption{ {\it Persistence as a function of channel width for different adhesion strengths}. (a) Vimentin-null cell (b) Wild-type cell.} 
  \end{figure*}

\begin{figure*}[hbt!]
\includegraphics{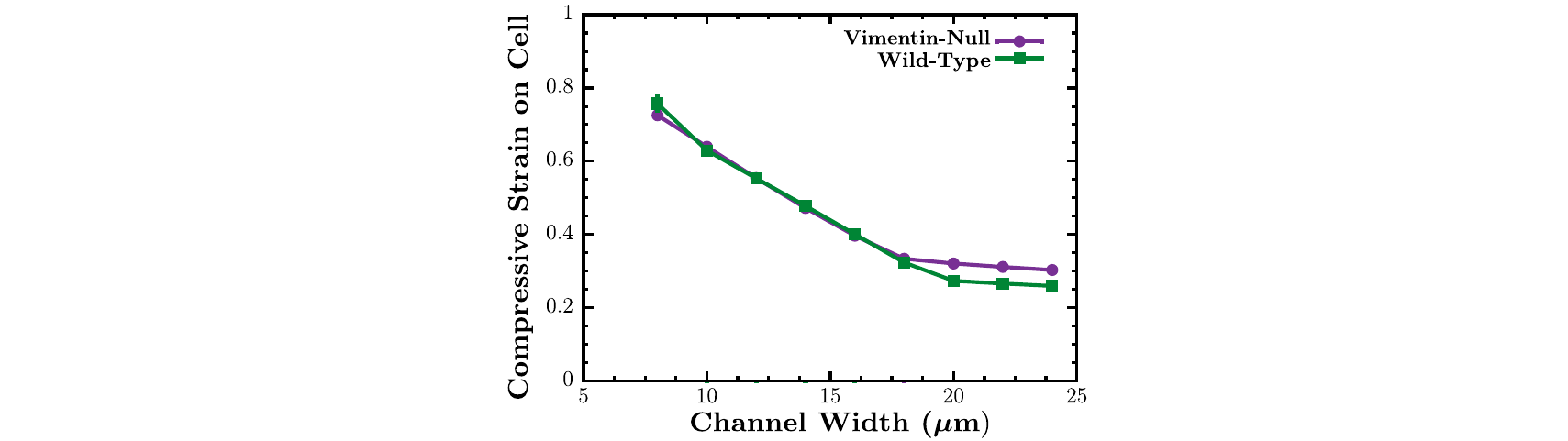}
\caption{{\it Strain on cell}. Compressive strain on the cell versus channel width for both cell types. } 
  \end{figure*}

\begin{figure*}[hbt!]
\includegraphics{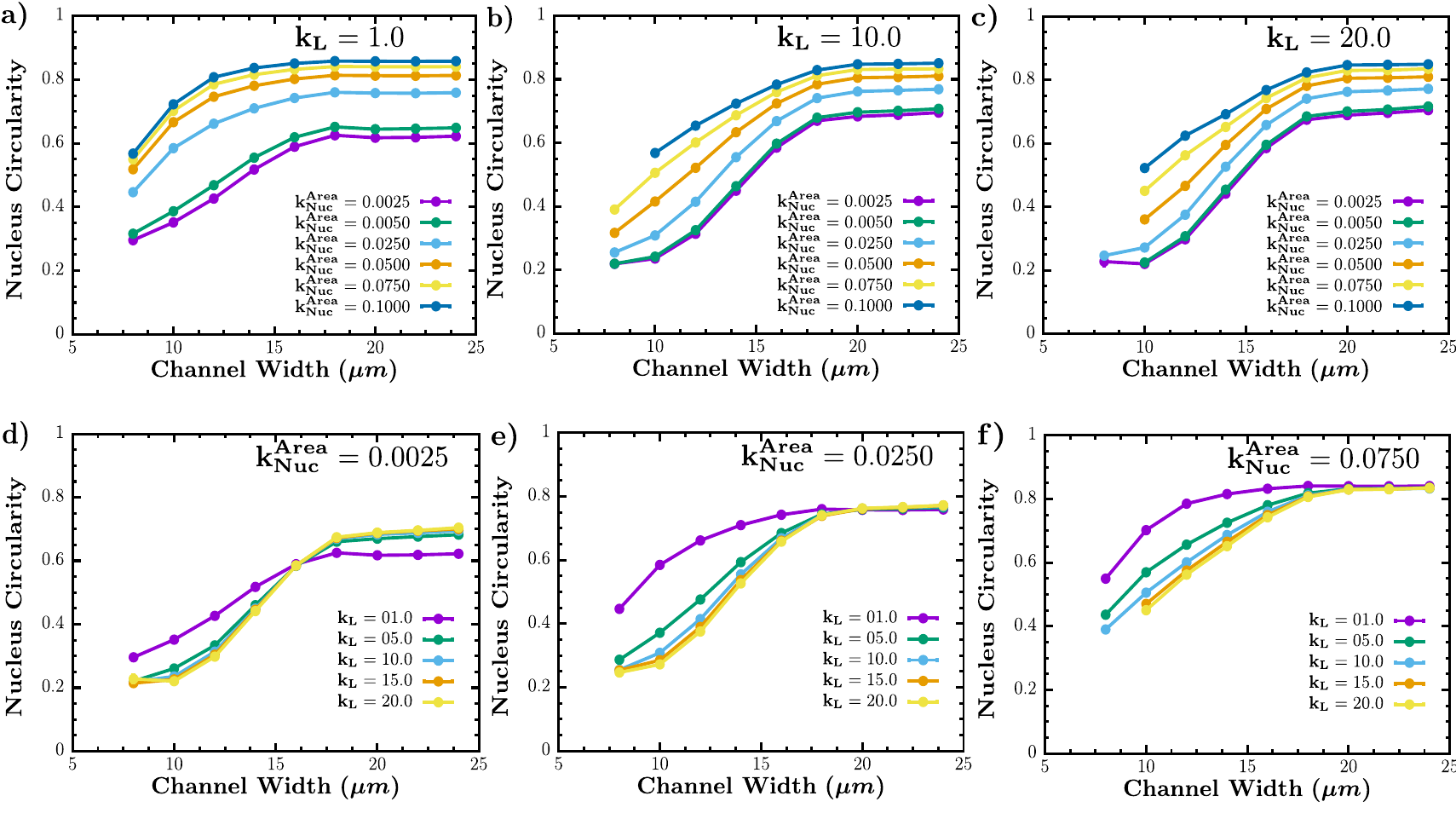}
\caption{{\it Nucleus circularity as a function of channel width for different linker and nuclear area spring strengths}: (a)-(c) Varying nuclear area spring strength $K_{nuc}^{area}$ for different linker spring strengths, $K_L$. (d)-(f) Varying linker spring strengths, $K_L$, for different nuclear area spring strengths  $K_{nuc}^{area}$.} 
  \end{figure*}

\end{document}